\documentclass[twocolumn]{revtex4}
\usepackage{amssymb,amsfonts,amsmath,graphicx,natbib}

\begin{document}



\title{Where are the missing baryons in clusters?}





\author{Bilhuda Rasheed, Neta Bahcall, and Paul Bode}

\affiliation{Department of Astrophysical Sciences,
4 Ivy Lane, Peyton Hall, Princeton University, Princeton, NJ 08544}





\begin{abstract}

Observations of clusters of galaxies suggest that they contain
significantly fewer baryons (gas plus stars) than the cosmic baryon fraction. This `missing baryon' puzzle is especially surprising
for the most massive clusters which are expected to be representative
of the cosmic matter content of the universe (baryons and dark
matter).  Here we show that the baryons may not actually be missing from clusters, but rather are extended to larger radii than
typically observed. The baryon deficiency is typically observed in the
central regions of clusters ($\sim$0.5 the virial radius). However,
the observed gas-density profile is significantly shallower than
the mass-density profile, implying that the gas is more extended
than the mass and that the gas fraction increases with radius. We use the
observed density profiles of gas and mass in clusters to extrapolate
the measured baryon fraction as a function of radius and as a function
of cluster mass.  We find that the baryon fraction reaches the cosmic
value near the virial radius for all groups and clusters above
$\sim5\times10^{13}h^{-1}_{72}M_\odot$. This suggests that the baryons
are not missing, they are simply located in cluster outskirts. 
Heating processes (shock-heating
of the intracluster gas, plus supernovae and AGN feedback)
that cause the gas to expand are likely explanations for these results. Upcoming observations should be able to detect these baryons.

\end{abstract}

\maketitle







\section{Introduction}

Clusters of galaxies, the largest virialized systems in the universe, are
powerful tools in constraining cosmology and tracing the large-scale
structure of the universe \cite[][and references therein]{bahcall88,
bahcall99, rosati, peebles}. The large mass of clusters ($\sim10^{14}$
to $10^{15}h^{-1}_{72}M_\odot$) implies that their contents---dark and baryonic
matter---have been accreted from very large regions of $\sim$10 comoving
Mpc, and therefore should be representative of the mean matter content
of the universe.  The strong gravitational potential of clusters also
implies that baryons cannot easily escape from these systems.  Therefore,
clusters are expected to retain the cosmic baryon fraction, i.e., the
relative fraction of baryons to total matter. This basic assumption
of a representative baryon fraction in clusters was used in 1993 \cite{white} to suggest that the mass-density of the universe must
be low, since the observed baryon fraction in clusters was considerably
larger than expected for a critical density universe.  Most of the
baryons in clusters reside in the X-ray emitting hot intracluster gas,
which approximately traces the cluster gravitational potential dominated
by dark matter.  The rest of the baryons are in the luminous galaxies
and in isolated stars that comprise the small amount of faint diffuse intracluster
light (ICL).

A puzzle has been raised, however, over the last few years: Detailed
X-ray observations from \emph{Chandra}, \emph{XMM-Newton} and others
suggest that the cluster baryon fraction (gas plus stars relative to
total mass) is considerably lower than the cosmic value \cite{afshordi,
umetsu, V06, arnaud, sun, giodini}. The cosmic baryon fraction is well
determined both from Big-Bang nucleosynthesis \cite{walker, burles}
and from observations of the Cosmic Microwave Background to be $f_b =
0.1675 \pm 0.006$ \cite[WMAP7:][]{jarosik}.  The cluster gas fraction
has been reported by the above observations to be only $60-80$\% of
the cosmic value, with stars contributing only a small ($\sim$10\%)
additional amount of baryons.  This baryon discrepancy is observed to
increase with decreasing cluster mass \cite{sun, giodini}. This raises the questions: Where are
the missing baryons? Why are they `missing'?

Attempted explanations for the missing baryons in clusters range from
preheating or other energy inputs which expel gas from the system
\cite[][and references therein]{cavaliere, metzler, takizawa, bialek,
valdarnini, mccarthy, bode}, to the 
suggestion of additional baryonic components not yet detected \cite[e.g.,
cool gas,  faint stars][]{bonamente, afshordi}.

In this paper we investigate the possibility that the `missing baryons'
are not missing at all, but are rather located in the outskirts of
clusters where few detailed observations have yet been made. The
`missing baryons' problem is typically observed within the central
regions of clusters, generally within a radius of $R_{500}$ (where the
enclosed mass-density is 500 times the critical density). This radius
is $\sim$0.5 of the virial radius of the cluster (where the enclosed
density is $\sim$100 times the critical density, for the current LCDM
cosmology \cite{eke, bryan}). Thus for a  virial radius of $\sim$1.5 Mpc,
the typical missing baryon problem is observed only at $\sim$0.75 Mpc from
the cluster center. 

Observations show that the gas density profile in
the outer parts of clusters decreases with radius slower than the mass
profile in these regions. 
Using gravitational lensing, the latter has been observed out
to large radii \cite{mandelbaum, sheldon, umetsu} and is consistent with the expected NFW profile
\cite{nfw}.  While the cluster mass-density declines with radius
approximately as $r^{-2.6}$ in these outer regions, the gas-density
declines only as $r^{-2.2}$. This implies that the gas is more
extended than the total mass, and therefore the gas fraction increases
with radius beyond the observed radius of $R_{500}$. A shallow slope of
the gas profile (as compared with the mass profile) is indeed expected
if gas heating occurs in the clusters (e.g., from shock-heating of the gas,
supernovae, and AGNs; these processes are known to exist). The heating
expands the gas relative to the dark matter potential and spreads it
out to larger radii, with a shallower slope, as observed.

We use the observed slopes of the gas-density and mass-density profiles in
the outer regions of clusters to extrapolate the observed gas fraction
from $R_{500}$ to larger radii, up to the virial radius ($R_{vir} = R_{100}$ \cite{eke,bryan}). We add the observed stellar fraction to the extrapolated gas fraction to find the baryon fraction at large radii.
We perform this extrapolation as a function of cluster mass, from groups
to rich clusters. Our analysis is based entirely on observations.

We find that there is no `missing baryon' problem in rich clusters when
the data is extrapolated to near the virial radius; at that radius,
the baryon fraction is consistent with the cosmic value.  Most of the
`missing baryons' are therefore expected to be in the outskirts of
clusters, between $R_{500}$ and $R_{vir}$. This result can be tested with
upcoming observations of the Sunyaev-Zeldovich effect in clusters (e.g.,
SPT \cite{plagge}; ACT \cite{fowler}) as well as with more sensitive
X-ray observations.

Observations have shown that the `missing baryon' problem at $R_{500}$ becomes
significantly more severe for lower mass clusters and groups of galaxies than for rich clusters;
the observed gas fraction decreases considerably with decreasing cluster
mass. This too would be expected if the heating processes expand
the gas: the lower gravitational potential of the smaller systems
will not be able to hold on to their gas as well as the higher mass
clusters. The gas density profile in small groups is indeed observed to be
shallower than the gas density profile in massive clusters, suggesting
that the gas in low mass systems is more spread out. We extrapolate the observed baryon fraction as described above
for clusters as a function of their mass---from groups to rich clusters.
We find that for this entire mass range the baryon fraction within $R_{vir}$ is flat and is consistent with the cosmic value.

In the next section we describe the observations and analysis. A discussion of the results and our conclusions follow. We use a LCDM cosmology with $h$=0.72 and $\Omega_m$=0.258.

\section{Observations, Analysis and Results}\label{obs}

The gas fraction in clusters of galaxies has been measured for
a relatively large sample of groups and clusters within $R_{500}$. The total baryon fraction is obtained by adding the stellar mass
fraction observed for these systems within the same radius. This baryon
fraction is systematically lower than the cosmic value measured by WMAP7;
the discrepancy increases with
decreasing cluster mass \cite[][and references therein]{sun,giodini}. Here we
investigate the possibility that the missing baryons are not actually
missing from clusters but are rather spread out to larger radii, beyond
$R_{500}$.
We investigate this scenario by extrapolating the observed gas fraction
to larger radii, from $R_{500}$ up to the virial radius ($R_{vir} = R_{100}$),
using the mean observed gas and mass density profiles in clusters;
these density profiles have been measured up to the virial radius (and
occasionally beyond). The observed stellar mass fraction, including the small contribution from the ICL, is added to the gas fraction to yield the total baryon fraction. The baryon fraction is then
investigated as a function of radius, from $R_{500}$ to $R_{vir}$,
and as a function of mass, from groups to rich clusters. 

\subsection{Gas Fraction at $R_{500}$}\label{fg}

Although some observations extend to $R_{200}$,
the gas fraction has been 
accurately measured for a sufficiently large sample of nearby clusters
only out to $R_{500}$.

We use X-ray observations of the gas fractions for 39 nearby clusters
observed with \emph{Chandra} and \emph{XMM-Newton} \cite{V06,
arnaud, sun}. These authors use similar methods of data reduction
and analysis. We use the compilation by Giodini et al.~\cite{giodini} of these groups and clusters above
the mass of $M_{500} = 3\times10^{13}h^{-1}_{72}M_\odot$ (where $M_{500}$ is the mass within $R_{500}$). After conversion to a common cosmology (LCDM with
$\Omega_m = 0.258$  and $H_0 = 72$ km/s/Mpc), the three samples
have been binned into four logarithmically spaced mass bins,
from groups to rich clusters \cite{giodini}. (We do not include
the lowest mass bin at $10^{13}h^{-1}_{72}M_\odot$ which
contains only two groups with large error bars.) Our cluster
sample has a mass range of $3\times10^{13}h^{-1}_{72}M_\odot <
M_{500} < 10^{15}h^{-1}_{72}M_\odot$ and a redshift range of
$0.012<z<0.23$. The mean observed gas fraction for each mass bin is
listed in Table 1 and shown in Figure 1. The error on the mean is the rms
standard deviation divided by $\sqrt{N-1}$. The horizontal bars are the mass ranges for the bins. Also presented in Figure 1
is the cosmic baryon fraction observed by the WMAP7 microwave background
measurements, $f_b = 0.1675 \pm 0.006$ \cite{jarosik}.  One can see that the cluster
gas fraction at $R_{500}$ is significantly lower than the cosmic baryon
fraction. The gas fraction decreases significantly from rich to poor clusters; while rich clusters contain about 12\%
gas within $R_{500}$, the gas fraction in poor clusters and groups is
only $\sim$6-7\%.

\subsection{Stellar Fraction}\label{fs}

The galactic stellar fraction has been measured in nearby clusters using
multiband optical and infrared surveys combined with stellar population
models. We use the results obtained from the COSMOS survey \cite{giodini}
and 2MASS survey \cite{lin} for nearby clusters. The combined COSMOS
and 2MASS sample covers our entire cluster mass range, 
$3\times10^{13} - 10^{15}h^{-1}_{72}M_\odot$. We bin the observed
stellar fraction into the same
four logarithmic mass bins as the gas fraction sample.
The mean stellar
fraction declines with cluster mass as $M^{-0.37\pm0.04}$
\cite[see][Figure 5]{giodini},
exhibiting an opposite trend to that of the gas fraction.

To determine the contribution of the diffuse intracluster light (ICL) to
the stellar fraction, we use the observations by Zibetti \cite{zibetti}
which use the Sloan Digital Sky Survey data to reach to unprecedented
cluster-centric distances and depth for a large number of stacked clusters
of various masses. They find the ICL is centrally concentrated,
and that on average the ICL contributes $\sim10$\%
of the stellar light within the central 500 kpc for all cluster masses.
We add this 10\% contribution to
the galactic stellar fraction discussed above for all clusters. We note
that the ICL contribution may decline to less than 10\% when extending to
larger cluster radii; but since the ICL is a very small contribution to
the total baryon fraction, this effect has negligible consequences (see Discussion). The
total stellar fraction for the four mass bins is summarized in Table
1. It is added to the gas fraction to obtain the total average baryon
fraction for each mass range. The baryon fraction within $R_{500}$ is
listed in Table 1 and plotted as a function of mass in Figure 1. The
deficiency of baryons within $R_{500}$ relative to the cosmic value is
clearly seen in Figure 1; the deficiency becomes more severe for lower
mass clusters.

\begin{table}
\caption{Cluster gas, stellar, and total baryon fractions within $R_{500}$}
\begin{tabular}{@{\vrule height 10.5pt depth4pt  width0pt}| c | c | c |
c | c |}
\hline
\hline
Bin&$<M_{500}>$&$f^{gas}_{500}$&$f^{stars+ICL}_{500}$&$f^{b}_{500}$\\
&$(h_{72}^{-1}M_\odot)$&&&\\[5pt]
\hline
\hline
1
&$5.1\times10^{13}$&$0.068\pm0.005$&$0.050\pm0.002$&$0.118\pm0.005$\\[5pt]
2
&$1.2\times10^{14}$&$0.080\pm0.003$&$0.040\pm0.004$&$0.120\pm0.005$\\[5pt]
3
&$3.0\times10^{14}$&$0.103\pm0.008$&$0.023\pm0.002$&$0.126\pm0.009$\\[5pt]
4
&$7.1\times10^{14}$&$0.123\pm0.007$&$0.021\pm0.002$&$0.143\pm0.007$\\[5pt]
\hline
\end{tabular}
\label{data}
Gas, stellar (including ICL), and baryon fraction of clusters within $R_{500}$ for four cluster mass bins: Bin 1 ($0.3-0.7\times10^{14}h^{-1}_{72}M_{\odot}$), Bin 2 ($0.7-1.7\times10^{14}h^{-1}_{72}M_\odot$), Bin 3 ($1.7-4.2\times10^{14}h^{-1}_{72}M_\odot$) and Bin 4 ($4.2-10\times10^{14}h^{-1}_{72}M_\odot$). $f^{gas}_{500}$ are averages from Chandra and XMM observations \cite{V06, arnaud, sun, giodini}. $f^{stars}$ and the 10\% ICL contribution are from \cite{giodini, lin, zibetti}. The error bars are 1-$\sigma$ errors on the mean.
\end{table}

\subsection{Gas and Mass Density Profiles}\label{slope}

Extrapolating the observed gas fraction to larger radii beyond $R_{500}$
requires the knowledge of the observed gas and mass density
profiles in these regions. The gas density profile has been measured
well in the outer parts of clusters ($R_{500} - R_{vir}$) using X-ray observations
of nearby clusters with \emph{ROSAT, Chandra, XMM-Newton},
and \emph{Suzaku}. The observed gas
profile in the outer regions fits well to a beta-model with
a density slope of $3\beta_{gas} = \alpha_{gas}$ (where $\rho_{gas}
\propto r^{-\alpha_{gas}}$). We use observations with small
uncertainties on the gas density slope at large radii and which cover our entire cluster mass range \cite{V06,
V99, dai, bautz, pratt}. 

The averaged observed $\beta$-values of the gas density slope are presented
as a function of cluster temperature in Figure 2. The data include measurements for 51 clusters as well as average slopes for stacked samples of hundreds of optical clusters; the slopes are binned in temperature as presented in Figure 2. The error bars on the bin-averaged
gas slopes are the standard deviation divided by $\sqrt{N-1}$. The
\emph{ROSAT} observations of 39 $z<0.25$ clusters by Vikhlinin et al.~\cite{V99} cover the full outer regions of clusters, from $0.3R_{500}$
to $1.5R_{180}$ ($\sim{R}_{vir}$). The best-fit slope
to these outer regions is determined for a wide range of cluster temperature, from 2
keV to over 10 keV. The \emph{Chandra} observations of 10 massive clusters by
Vikhlinin et al.~\cite{V06} provide $\beta$-fits to gas density slopes
near $R_{500}$, nicely consistent with the trend shown by the slopes of
the previous sample \cite{V99}. For the lowest mass bin ($M_{500} \approx
5\times10^{13}h^{-1}_{72}M_\odot$) we use the observed density slope
from \emph{ROSAT} by Dai et al.~\cite[][their richness class 1]{dai}, who obtain integrated X-ray gas profile for stacked samples of hundreds of low-mass optical clusters out to $R_{vir}$. We also include Bautz et al.~\cite{bautz} who measure the gas profile of Abell
1745 to $R_{200}$ using \emph{Suzaku} observations. 

The observed beta slopes presented in Figure 2 are consistent with each other, and show a shallower slope for lower mass
systems than for massive clusters. For our four mass bins (Table 1,
Figure 1) we find the following mean gas density slopes: $\alpha_{gas}
= 3 \times \beta_{gas} = 1.8\pm0.2$ for Bin 1, $1.9\pm0.07$ for Bin 2,
$2.1\pm0.02$ for Bin 3, and $2.3\pm0.02$ for Bin 4. The error bars are
the standard deviation of the average $\beta$-values in each bin divided
by $\sqrt{N-1}$.  When comparing a slope at a given temperature bin in
Figure 2 to a given mass bin in Figure 1, we use the observed $M_{500}-T$
relation \cite{V06}; the results are not sensitive to the exact conversion
because of the slowly varying $\beta_{gas}(T)$ relation.


The final piece required for extrapolating the gas (and baryon) fraction to large radii is the total mass density profile. This has been observed to follow the
NFW profile \cite{nfw} to large radii, by using weak lensing observations
with the Sloan Digital Sky Survey and other observations \cite[][and
references therein]{mandelbaum, sheldon, umetsu}. We use the average
observed value of the concentration parameter $c_{200} = 5$ for our
mass range \cite{mandelbaum}. The NFW profile has a mass density slope
of $\alpha_m = 2.6$ in the radius range from $R_{500}$ to $R_{200}$,
and $\alpha_m = 2.7$ from $R_{200}$ to $R_{vir}$ (where $\rho_m \propto
r^{-\alpha_m}$). These slopes are considerably steeper than the
corresponding gas density slopes, thus yielding an
increasing gas fraction with radius in cluster outskirts.

\subsection{Extrapolation and Results}

Using the observed gas density and mass density slopes at large radii,
we extrapolate the observed gas fraction from $R_{500}$ as a function of
radius up to the virial radius. The gas fraction increases with radius as:

\begin{equation*}
f^{gas} (< R) \propto \frac{\rho_{gas}(R) \propto
R^{-\alpha_{gas}}}{\rho_m(R) \propto R^{-\alpha_m}} \propto R^{\alpha_m -
\alpha_{gas}}, \hspace{0.7cm} r>R_{500}.
\end{equation*}

The extrapolated gas fraction from $R_{500}$ to $R_{200}$ and to $R_{vir}$
is presented as a function of cluster mass in Figure 3a. Because the
gas density slope is shallower in groups than in rich clusters, the
increasing trend of gas fraction with mass becomes weaker
at the outer radii.

The baryon fraction is presented for the different radii---$R_{500}$,
$R_{200}$, and $R_{vir}$---as a function of mass in Figure 3b; the baryon
fraction within these radii is the sum of the gas fraction (Figure
3a) and the stellar mass fraction discussed above, including the 10\%
ICL. We assume that this fraction
remains constant with increasing radius; the main results do not change
significantly if this assumption is changed because of the relatively
small contribution of the stellar component (see Discussion). The error
bar on the extrapolated gas fraction is the propagated errors of the
gas fraction at $R_{500}$ and the gas slope used for extrapolation. The
error bar on the baryon fraction is the combined errors of the gas and stellar fractions.

The results in Figure 3b show that the baryon fraction flattens
considerably as a function of cluster mass when extrapolated to larger
radii; this is due to the combined effect of the shallower gas density
profile in groups, which results in more gas in their outskirts,
plus the larger observed stellar mass fraction in groups, which
adds more baryons in the smaller systems. In fact, at the virial
radius we find that the baryon fraction is essentially flat from groups to rich clusters, at a level consistent with the
cosmic baryon fraction. This suggests that there are no
`missing baryons': most of the `missing baryons' are likely located in
the outskirts of clusters, extending to nearly the virial radius.

The extrapolated baryon fraction is presented as a function of radius
for two of our mass bins (Bins 2 and 4) in Figure 4; the results show
the slow but steady increase in the baryon fraction with radius.

\section{Discussion}

The gas fraction at $R_{500}$ is often cited to claim that
clusters contain significantly fewer baryons than the universal baryon fraction and
therefore exhibit a `missing baryon' problem. Here we show, based purely
on observational results, that the gas (and baryon) fraction increases substantially
with radius beyond $R_{500}$; the radius $R_{500}$
cannot be used as a representative radius for comparisons with the
global value. Using the observed gas and mass density profiles, we extrapolate the observed baryon fraction (gas and stars)
as a function of radius from $R_{500}$ to the virial radius. Since the
gas-density is observed to decline more slowly with radius than the
total mass, we find that the average baryon fraction increases with radius for clusters of all
masses; it reaches the cosmic baryon
fraction near the virial radius ($R_{100}$). This suggests that baryons
are not missing in clusters, they are simply located in cluster outskirts.

Recent observations of the Sunyaev-Zeldovich (SZ) effect in 15 massive
clusters using the South Pole Telescope \cite{plagge} measure the
gas density pressure profile in these clusters as a function of radius up to the
virial radius (and beyond); they are well fit with a beta-model in the
outer regions. The detection of the gas to these large radii, and their
observed beta-model slope (when corrected for the temperature profile),
are nicely consistent with our results and the conclusion that the baryons are out there between $R_{500}$ and $R_{vir}$.
Similarly, George et al.~\cite{george} use \emph{Suzaku} X-ray observations to
trace the gas density profile in the massive cluster PKS0745-191 up to
the virial radius, observing the baryons at the cluster outskirts and
measuring a shallow gas-density profile in these outer regions.

The gas density profile is observed to be even shallower in low-mass
clusters than in massive clusters. This is consistent with the gas being heated via shocks and feedback from supernovae and AGN. This feedback will be more significant in low-mass systems when compared to the binding energy of the gas. If star formation is more efficient in groups than in clusters
\cite[e.g.,][]{lin}, this will further increase the gas entropy in these
systems, because the star formation removes the lowest-entropy gas from the
intracluster medium, leaving behind gas with higher average entropy
\cite{VoitBryan01}. A higher stellar fraction also implies more supernovae/AGN
activity per unit mass. This explains why, at $R_{500}$, groups have
a lower baryon fraction than clusters. However, by the virial radius,
we are able to account for all the baryons expected
from the cosmic value.

A few comments on uncertainties are in order. First, the gas
slopes we use from \emph{ROSAT} observations \cite{V99, dai}
are the best-fit slopes in the large radial range from $R_{500}$
($\sim{0.5}R_{vir}$) to $\sim{R}_{vir}$. This is the average slope in
these outer cluster regions which  are relevant to our extrapolation. If
we instead use the best measured gas-density slope at $R_{500}$
\cite[e.g.,][]{V06} and then steepen it slowly with radius similar to
the steepening trend observed for the NFW profile (which
steepens by 10\% from $R_{500}$ to $R_{vir}$), we find
consistent results within the error-bars with the ones presented above.

The stellar fraction has been observed for a large number of clusters
by Gonzalez et al.~\cite{gonzalez}, who report a somewhat higher stellar fraction for
low-mass clusters and a slightly lower stellar fraction for high-mass
clusters than the observations by Giodini et al.~\cite{giodini} and Lin et al.~\cite{lin}. The higher stellar fraction in groups \cite{gonzalez} is likely due to a selection bias towards systems with dominant Brightest Cluster Galaxies in the small groups. Their observed
trend of stellar fraction with mass is therefore somewhat steeper,
${\log}f^{stars} = (7.57\pm0.08) - (0.64\pm0.13)\log{M}_{500}$. Using
this stellar fraction in our analysis does not change our result significantly; the baryon fraction at the virial radius decreases by $\approx5\%$ for the most massive bin, and increases by $\approx7\%$ for the lowest mass bin, but all are well
within our 1-$\sigma$ error bars.

We use the observed 10\% contribution of the ICL to the stellar fraction \cite{zibetti}. If the ICL fraction changes
at the outer radii, it will not affect our results
since the ICL contribution is only $\sim$0.2--0.4\% of total mass.

Similarly, we assume in our analysis that the stellar fraction does not
change when extrapolated from $R_{500}$ to $R_{vir}$, since no observations
of the stellar fraction have been made beyond $R_{500}$. If the stellar
mass fraction decreases somewhat with radius, then the baryon fraction
will slightly decrease. If we assume, for example, a 20\% decrease in
stellar fraction from $R_{500}$ to $R_{vir}$, we find that the baryon
fraction is lowered by only 2\% for the massive clusters and 8\% for the
low-mass groups (at $R_{vir}$); these values are within 1-$\sigma$
of our baryon fraction results even for the lowest mass groups.

For the mass density profile, we use the mean observed value of the
concentration parameter of the NFW profile, $c_{200} =5$, as observed from
weak lensing \cite{mandelbaum} for our mass range. As Mandelbaum et al.~\cite{mandelbaum} discuss in their
paper, this value is slightly lower than results from simulations and
some previous studies. If $c_{200}$ is larger than 5, the mass
profile will be more concentrated, i.e., fall off even steeper with
respect to the gas profile at the outer radii. This would cause
the gas fraction to increase even more with radius. The effect is
small, however, and a change of $c_{200}$ to 7 induces a change in baryon
fraction at the virial radius of $<$5\%. Earlier in this Section we
discussed a scenario in which the gas density declines at the same trend
as the NFW profile; in that scenario, the effects will nearly cancel out.

The observations presented above are consistent with an energy input
in clusters that heats the intracluster gas and expands it to larger radii,
where the missing baryons may be found. Hydrodynamic simulations of cluster formation
which include the relevant physics are in qualitative good agreement (within $\sim$10\%)
with the picture presented here.  Using simulations without cooling, star formation,
or feedback, the baryon fraction is roughly constant, at 90\% of the cosmic value, from $R_{500}$ to $R_{200}$ \cite{Crain07}.   When these processes are included (with feedback coming from
AGN), then the baryon fraction is found to be lower at  $R_{500}$,  but instead of being flat it increases with radius \cite{Sijacki08,Battaglia10}.
Similar results, based on cluster energetics, are found in the models of Bode et al.~\cite{bode}.
Further improvements in models and simulations (including additional physical processes,  sources of nonthermal pressure and possible non-equipartition effects) will be needed to provide precise comparisons with the observations.

\section{Conclusions}

We investigate the `missing baryon' problem in clusters of galaxies.
Observations show that the baryon fraction (gas plus stars) measured
within a radius of $R_{500}$ in groups and clusters is significantly
below the cosmic value; this baryon discrepancy increases with decreasing
cluster mass. This gives rise to the puzzle: Where are the missing baryons? Why are they missing? We show that the baryons may not be missing at all, but rather are spread
out to larger radii, beyond $R_{500}$, and can be found in the cluster
outskirts. Here we show that the radius $R_{500}$, which is typically used in
observations, cannot serve as representative for the baryon fraction in
groups and clusters.  Based entirely on observations, we investigate
the dependence of the baryon fraction on radius for clusters of different masses, from groups to rich
clusters.  We use the mean observed gas and mass density profiles
in clusters to extrapolate the observed gas fraction as a function
of radius from $R_{500}$ to the virial radius. Since the gas-density
profile is significantly shallower than the mass-density profile, the
gas-fraction increases with radius; it increases more rapidly for groups than for rich clusters because of shallower gas-density
slope in groups. We add the observed stellar fraction and the diffuse intracluster
light  to the gas fraction to obtain the total baryon fraction. We find that the average baryon
fraction for all groups and clusters with $M_{500} \geq 5\times10^{13}h^{-1}_{72}M_\odot$
increases steadily with radius, reaching the cosmic value and becoming flat as a function of mass when measured within the virial radius (for the LCDM cosmology). This suggests that baryons are not missing in clusters,
but are simply located in cluster outskirts. This picture is consistent
with heating processes (such as shock-heating of the intracluster gas,
as well as supernovae and AGN feedback) causing the gas to expand to the cluster outskirts. Upcoming observations in the X-rays and SZ should be able to detect these baryons.





\begin{acknowledgments}
B.~R. owes a debt of gratitude to her research adviser, collaborator and friend Neta Bahcall for her judicious and delightfully rewarding stewardship of my senior thesis, which culminated in this paper. B.~R. warmly thanks Princeton University for enabling this experience. The authors thank Rachel Mandelbaum, Greg Novak, David Spergel, Michael Strauss and Alexey Vikhlinin for their helpful comments. Computational work was performed at the TIGRESS high performance
computer center at Princeton University, which is jointly supported by
the Princeton Institute for Computational Science and Engineering and
the Princeton University Office of Information Technology.
\end{acknowledgments}

\begin{figure}
\begin{center}
\includegraphics[bb=0in 4in 8.6in 9.5in, width=6.2in]{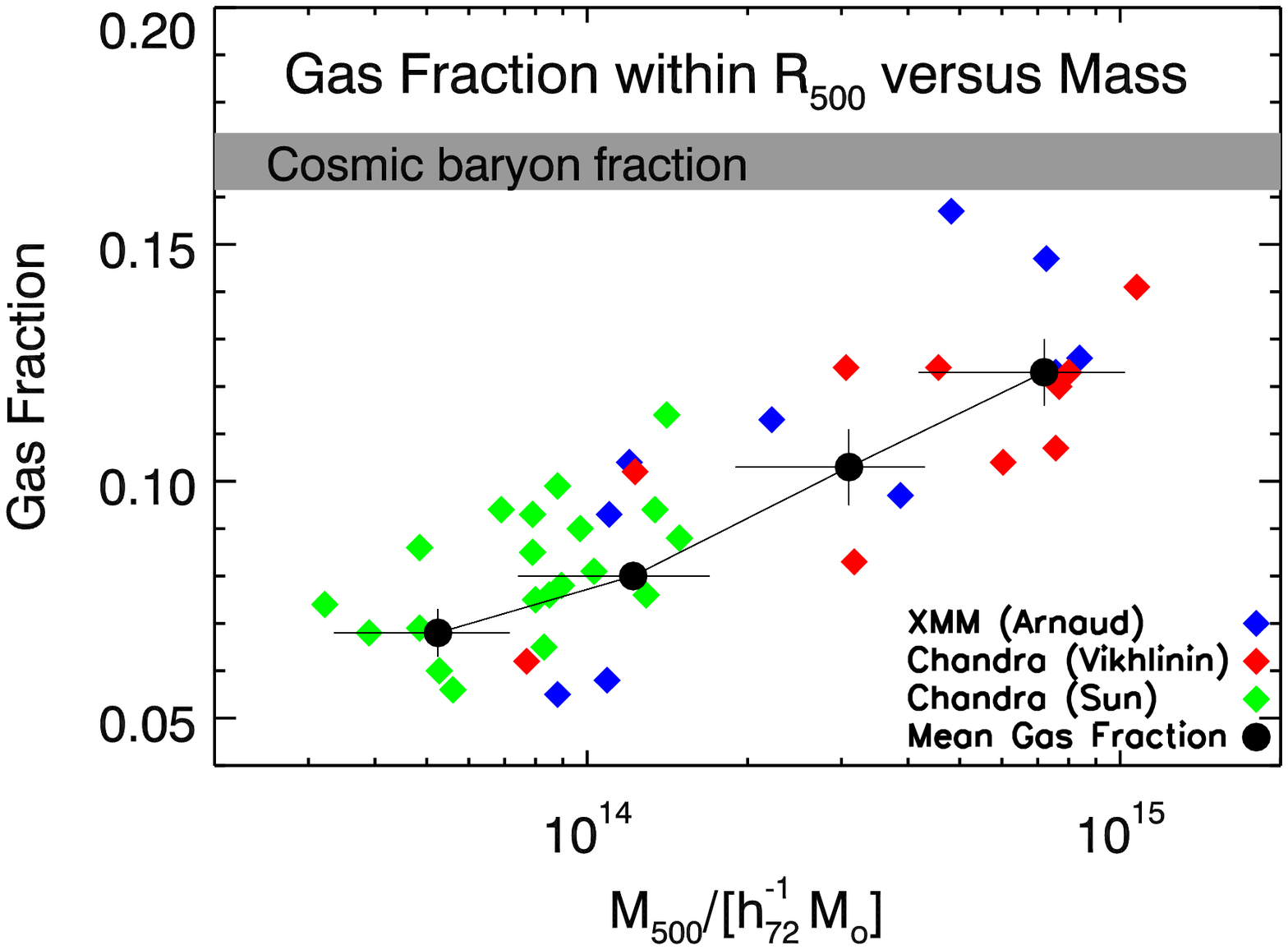}\label{newtrywgiod}
\includegraphics[bb=0in 4in 8.6in 9.5in, width=6.2in]{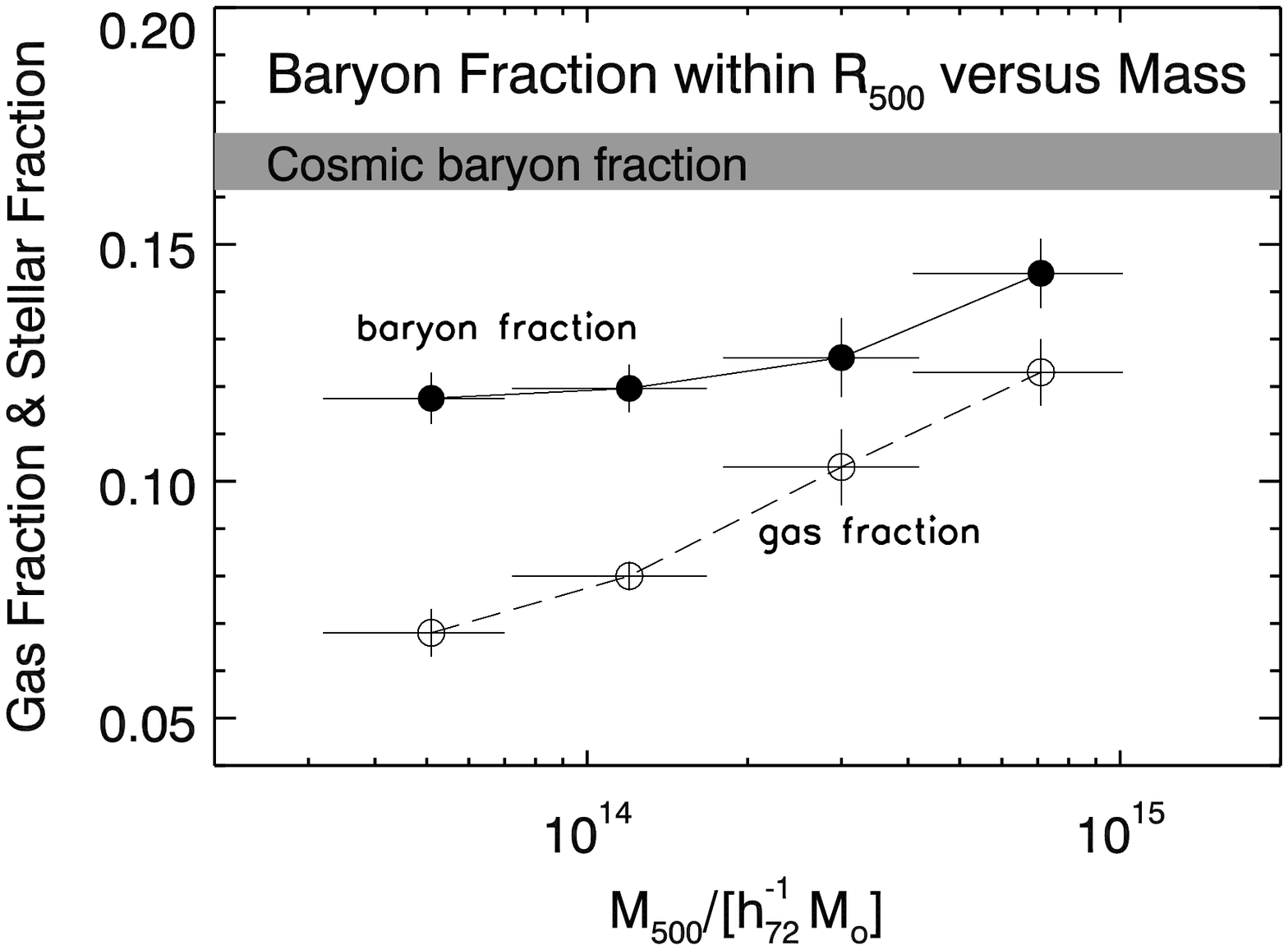}\label{baselineicl}
\caption{The observed cluster gas and baryon fraction within $R_{500}$ as a function of cluster mass, $M_{500}$. The observed cosmic baryon fraction is shown by the shaded band (1-$\sigma$; \cite{jarosik}). (a) Observed cluster gas fraction from \emph{Chandra} and \emph{XMM-Newton} within $R_{500}$ \cite{V06, arnaud, sun, giodini}. The average for the four mass bins is shown by the filled circles. (b) Observed cluster gas and baryon fraction within $R_{500}$ for the averages of the four mass bins. Error bars are 1-$\sigma$ errors on the mean. Horizontal bars represent the mass range of each bin.}
\end{center}
\end{figure}

\begin{figure}
\begin{center}
\includegraphics[bb=0in 4in 8.6in 9.5in, width=7.5in]{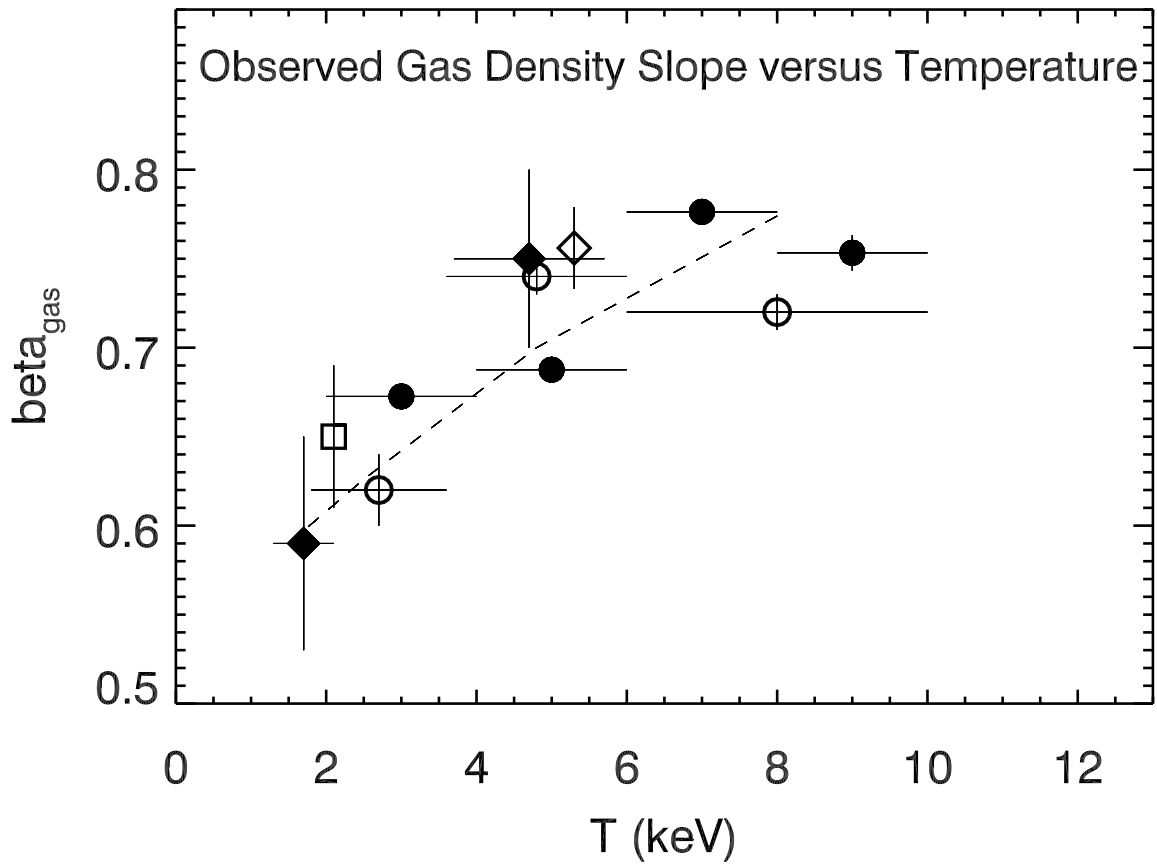}
\caption{Average observed gas density slope $\beta_{gas}$ (where $\rho_{gas} \propto r^{-3\beta_{gas}}$) at $\geq R_{500}$ from \emph{ROSAT, Chandra, XMM-Newton,} and \emph{Suzaku} \cite{V99, dai, V06, pratt, bautz} as a function of cluster temperature. Filled circles are bin-averages of 39 \emph{ROSAT} clusters \cite{V99}; filled diamonds are \emph{ROSAT} averages of hundreds of stacked optical clusters~\cite[][their richness classes 0 and 1]{dai}; empty circles are bin-averages of 10 \emph{Chandra} clusters \cite{V06}; empty diamond is \emph{Suzaku} observation for Abell 1795 \cite{bautz}; empty square is \emph{XMM} observation for A1983 \cite{pratt}. The dashed line illustrates the average trend of $\beta_{gas}$ with $T$. Error bars are 1-$\sigma$ errors on the mean. Horizontal bars represent the temperature bin size.}
\end{center}
\end{figure}

\begin{figure}
\begin{center}
\includegraphics[bb=0in 4in 8.6in 9.5in, width=6.2in]{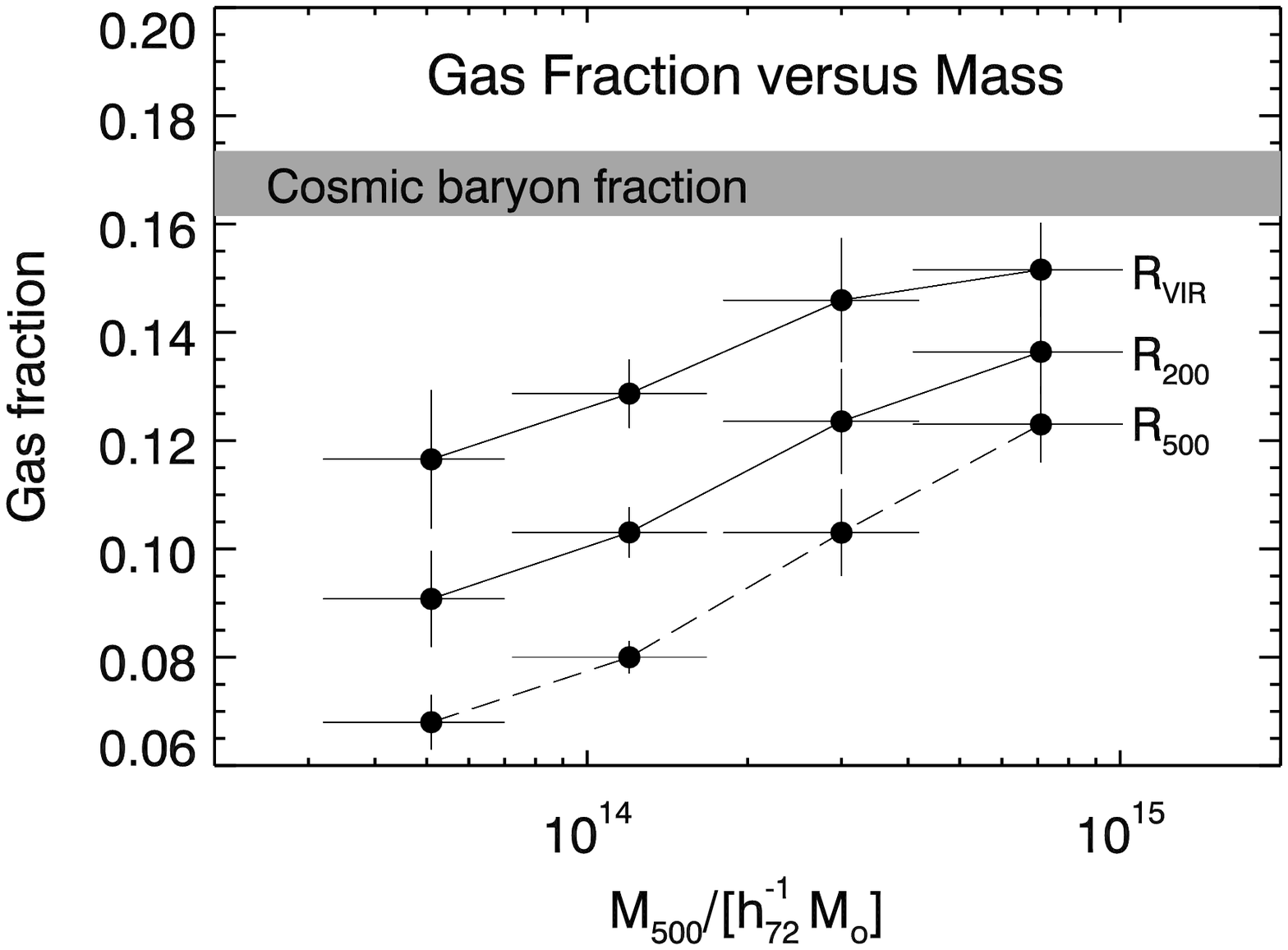}\label{giodgasextlow}
\includegraphics[bb=0in 4in 8.6in 9.5in, width=6.2in]{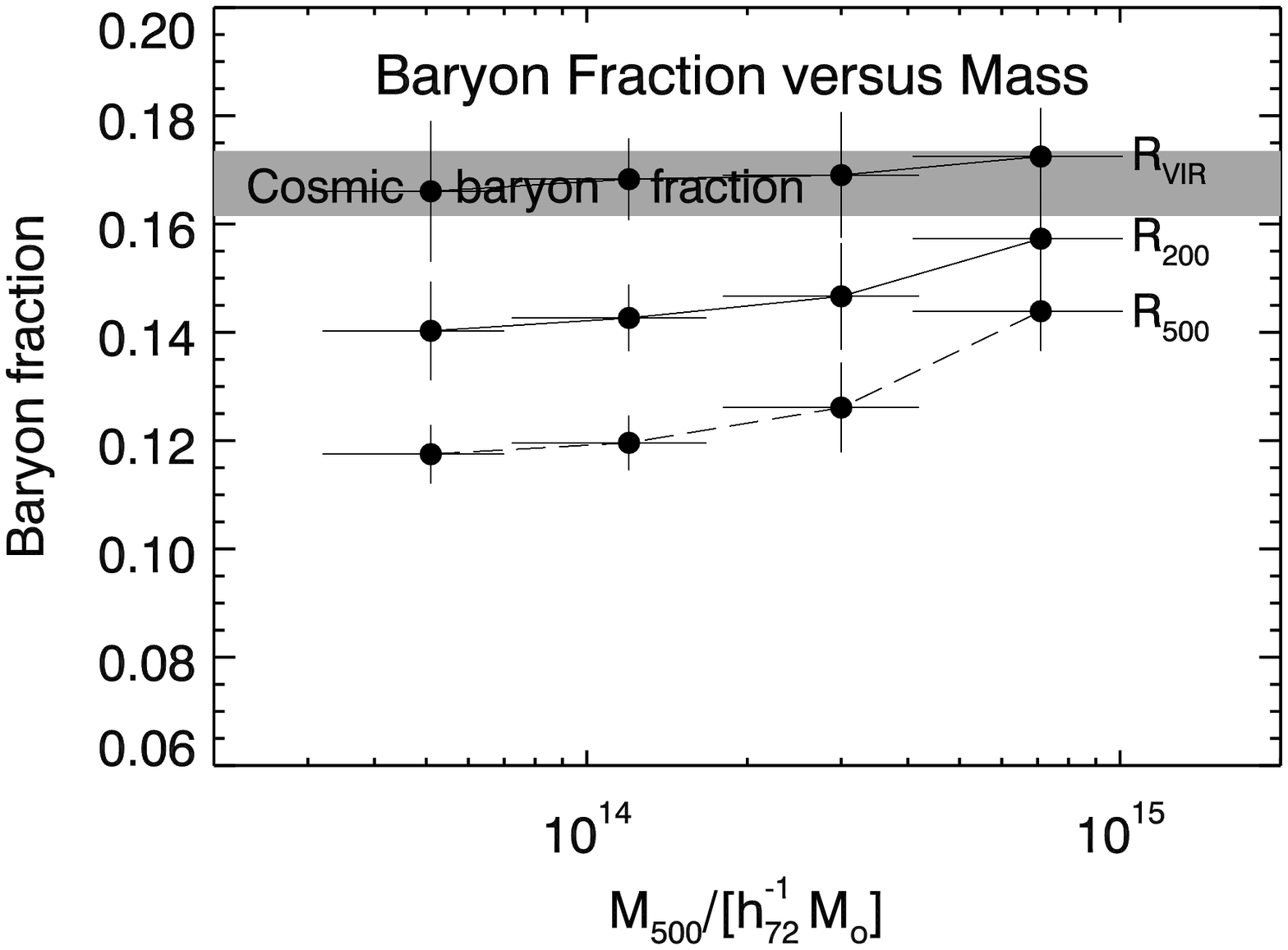}\label{giodbarextlow}
\caption{Cluster gas and baryon fraction (3a and 3b, respectively) within $R_{500}$ (observed),   $R_{200}$  and $R_{vir} (=R_{100})$ (extrapolated using observational data as discussed in the paper).  The  gas and baryon fractions are presented as a function of cluster mass for the four averaged binned samples. The error-bars are the 1-$\sigma$ error on the mean for each bin; horizontal bars represent the mass range of each bin. The observed cosmic baryon fraction is presented by the shaded band (1-$\sigma$).}
\end{center}
\end{figure}

\begin{figure}
\begin{center}
\includegraphics[bb=0in 4in 8.6in 9.5in, width=7.5in]{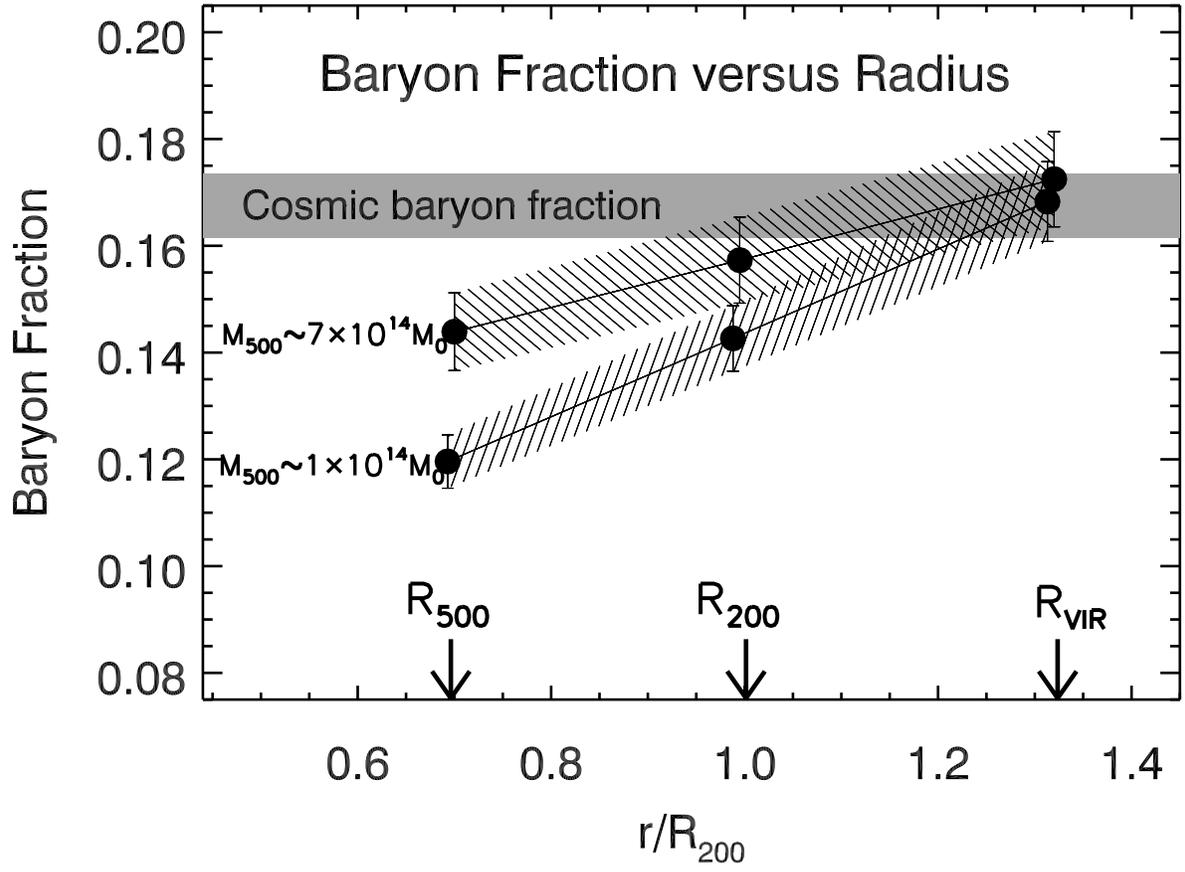}
\caption{The dependence of the baryon fraction on radius, from $R_{500}$ to $R_{vir} (=R_{100})$,  for two representative mass bins (bins 2 and 4; the others show a similar trend).  The radius is presented in units of $R_{200}$.  The slow but steady increase of the baryon fraction with radius is apparent, reaching the cosmic value near $R_{vir}$  for all cluster masses. The error-bars are 1-$\sigma$ errors on the mean.}
\end{center}
\end{figure}




\end{document}